\documentclass[prb,twocolumn,superscriptaddress,showpacs,floatfix]{revtex4}
\usepackage{graphicx,bm,amssymb}

\newcommand{\pdag}{{\phantom{\dagger}}}
\newcommand{\sign}{{\rm sgn}}

\begin{document}

\title{Quantum phase transition in a two-channel-Kondo quantum
dot device}

\author{M. Pustilnik} 
\affiliation{School of Physics, Georgia Institute of Technology, 
Atlanta, GA 30332}
\author{L. Borda}
\affiliation{Sektion Physik and Center for Nanoscience, LMU M\"unchen, 
Theresienstrasse 37, 80333 M\"unchen, Germany}
\affiliation{Hungarian Academy of Sciences, Institute of Physics, 
TU Budapest, H-1521, Hungary }
\author{L.I. Glazman} 
\affiliation{William I. Fine Theoretical Physics Institute, 
University of Minnesota, 
Minneapolis, MN 55455}
\author{J. von Delft}
\affiliation{Sektion Physik and Center for Nanoscience, LMU M\"unchen, 
Theresienstrasse 37, 80333 M\"unchen, Germany}


\begin{abstract}
  We develop a theory of electron transport in a double quantum dot
  device recently proposed in Ref.~\onlinecite{OGG} for the
  observation of the two-channel Kondo effect. Our theory provides a
  strategy for tuning the device to the non-Fermi-liquid fixed point,
  which is a quantum critical point in the space of device parameters.
  We explore the corresponding quantum phase transition, and make
  explicit predictions for behavior of the differential conductance in
  the vicinity of the quantum critical point.
\end{abstract}

\pacs{
72.15.Qm,    
73.23.-b,      
73.23.Hk,     
73.63.Kv      
}
\maketitle

\section{Introduction}
The magnetic screening of a localized spin by spins of itinerant
electrons~\cite{PWA_book} leads to the Kondo effect -- an anomaly in
low-temperature conduction properties. This screening becomes effective 
below some characteristic temperature, the Kondo temperature $T_K$. 
Above $T_K$ electrons are weakly scattered by the magnetic impurity, but below 
$T_K$ the scattering becomes strong. In the simplest Kondo systems, only one electron 
mode (the $s$-wave mode, say) participates in the screening of a localized spin 
with $S=1/2$. In this case, the low-temperature electronic properties are 
adequately described by Fermi liquid theory~\cite{Nozieres}, and the 
thermodynamic and transport characteristics are analytical functions 
of $T/T_K$. In more complicated systems (such as, e.g., paramagnetic 
metals) many electron modes may participate in screening of an $S=1/2$ 
localized moment~\cite{LM}. The peculiarities of such a ''multichannel'' 
Kondo model were long recognized~\cite{LM,NB}. At the same time it was 
understood that even a small deviation from symmetry between channels leads 
at low temperatures to the Kondo screening by just one channel, the one 
for which the exchange integral with the impurity is the largest~\cite{NB}.

The peculiarity of a {\it symmetric} multichannel Kondo problem is in
its non-Fermi-liquid (NFL) behavior at low temperatures~\cite{NB}. 
The low-temperature asymptotes of the thermodynamic and transport
characteristics display power-law behavior with fractional values of
the exponents. A complete temperature dependence of the thermodynamic
characteristics (such as the local spin susceptibility) is known now
from the exact Bethe-ansatz solution of the Kondo
problem~\cite{Bethe,AJ}. Details of the low-temperature
electron scattering problem were also understood in the framework of
 conformal field theory~\cite{Tsvelik,AL}.

Experimental observation of the non-Fermi-liquid behavior in a Kondo
system, however, is difficult because the channel symmetry is not
``protected'' -- in general, there are no conservation laws
prescribing such a symmetry. This has lead to various propositions to
observe such a behavior in systems where the role of spin is taken over 
by another degree of freedom, while the ``real'' spin labels the channels, 
making the channel symmetry robust. One such idea deals with an
atomic defect which occupies two equivalent lattice sites, thus
forming a pseudospin~\cite{CZ}. However, the equivalence of sites is
not a protected symmetry; its violation~\cite{Aleiner}, equivalent to
a ``Zeeman splitting'' of the pseudospin states, destroys the Kondo
effect.

Another object which under certain conditions can be described by the
two-channel Kondo model (2CK) model, is a large quantum dot, or a
metallic island connected by a single-mode channel to a conducting
electrode~\cite{Matveev}. If one neglects the finite level spacing in
the island, then a pseudospin labeling of the charge states of the
island may be introduced, while real spin again plays the part of the
channel index. In this setup the degeneracy with respect to the
pseudospin orientation is easily achieved by tuning the gate voltage
to the vicinity of the Coulomb blockade degeneracy point. At
temperatures $T$ higher than the level spacing $\delta E$ in the
island, the system is then described by the 2CK model~\cite{Matveev}.
Since $T_K$ for this system can be of the order~\cite{GHL} of the 
charging energy $E_C$, while typically $\delta E\ll E_C$, the NFL 
regime is easily realized. When an additional electrode is attached to 
the island, one can study the transport properties of the resulting
device. The disadvantage of such realization of a 2CK system is that
there is no mapping between the conductance across the
island~\cite{FM} and the electron scattering cross-section in the
generic two-channel Kondo model~\cite{Tsvelik,AL}.

Small quantum dots with large level spacing have proved to be suitable
for the observation of the Kondo effect~\cite{kondo_exp}. In the usual
geometry consisting of a dot with two attached electrodes, however,
only the conventional Fermi-liquid (FL) behavior is observable at low 
temperatures. The reason lies in the structure of the matrix of exchange 
constants that couple the dot's spin to the spins of itinerant electrons~\cite{real,tutorial}. 
Typically, the eigenvalues of this matrix are vastly different~\cite{real}, 
and their ratio is not tunable by conventional means.

\begin{figure}[h]
\includegraphics[width=0.55\columnwidth]{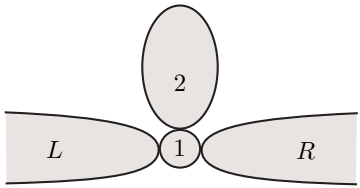}
\caption{Device proposed in Ref.~\onlinecite{OGG}. Level spacing in
the larger dot ($2$) must be negligibly small to allow for the NFL
behavior of the device at low temperatures.
\label{device}
}
\end{figure}

A device that circumvents this problem was proposed recently in 
Ref.~\onlinecite{OGG}, and involves several dots. A two-dot device 
is sufficient for the realization of the 2CK model. The key idea of 
Ref.~\onlinecite{OGG} is to replace one of the electrodes in the 
standard configuration by a very large quantum dot $2$, see 
Fig.~\ref{device}, characterized by a level spacing $\delta E_2$ and 
a charging energy $E_2$. At $T\gg \delta E_2$, particle-hole excitations 
within this dot are allowed, and electrons of dot $2$ participate in the 
screening of the smaller dot's spin. At the same time, as long as $T\ll E_2$, 
the number of electrons in the dot $2$ is fixed. As a result, the electrons 
in dot $2$ provide for a separate channel which does not mix with the 
channels provided by the electrodes $L$ and $R$. In this case, the 
exchange constants for two channels may be tuned to become 
equal~\cite{OGG}: the asymmetry between the channels is controlled 
by the ratio of the conductances of the dot-leads and dot-dot junctions.

In principle, a setup having just one lead and two dots would allow
one to study thermodynamic properties, such as magnetic susceptibility, 
in the 2CK regime. The existing technology~\cite{kondo_exp}, however, 
enables one to measure transport rather than thermodynamic properties. 
Therefore, two leads are needed to perform conductance measurements. 
In this paper, we assume that one of the electrodes is coupled weakly to 
the small dot and serves as a probe of the 2CK system formed by the 
two dots and the remaining electrode. We propose a detailed strategy for
tuning the device to the NFL regime, and discuss various manifestations 
of NFL-related physics in the transport properties of the system.

\section{The Model}
\label{Sec2}

According to the discussion above, the device we consider consists of two 
quantum dots coupled to two conducting leads via single-mode junctions. 
The model Hamiltonian of such a device can be written as a sum of three parts, 
\begin{equation}
H= H_d +H_l +H_t .
\label{2.1}
\end{equation} 
The first term here, $H_d$, describes an isolated system of two quantum 
dots, $1$ and $2$, connected via a single mode junction,
\begin{eqnarray}
H_d &=& 
E_1\left(\sum_s d^\dagger_s d^\pdag_s - N\right)^2
\nonumber \\
&+& \sum_{ks} \xi_k^\pdag \psi^\dagger_{2ks}\psi^\pdag_{2ks} 
+ E_2\left( \sum_{ks}\psi^\dagger_{2ks}\psi^\pdag_{2ks}\right)^2
\nonumber\\
&+& \sum_{ks} (t_2^\pdag \psi^\dagger_{2ks}d^\pdag_s + {\rm H.c.}
)
\label{2.2} 
\end{eqnarray} 
The last two terms in Eq.~(\ref{2.1}) represent the free electrons 
with spin $s = \pm 1$ in leads 
$R$ and $L$, and the tunneling between the leads and dot $1$, 
see Fig.~\ref{device},
\begin{eqnarray}
H_l &=& \sum_{\alpha ks} 
\xi_k^\pdag c^\dagger_{\alpha ks}c^\pdag_{\alpha ks},
\quad
\alpha=R,L\,;
\label{2.3} \\
H_t &=& \sum_{\alpha ks}t_{\alpha}^\pdag c^\dagger_{\alpha ks}d^\pdag_s 
+ {\rm H.c.}
\label{2.4}
\end{eqnarray} 
In Eq.~(\ref{2.2}) the smaller dot (dot $1$) is described by a
single-level system equivalent to the Anderson impurity model. The 
parameter $E_1$ represents charging energy, while the parameter $N$ is
adjustable by tuning the potential on the capacitively coupled gate
electrode. We neglect the finite level spacing $\delta E_2$ in the dot
$2$, but account for its finite charging energy $E_2$ (we do not write
explicitly the gate potential applied to the dot $2$, as it
corresponds to a trivial shift of the chemical potential).

Since the relevant energies ($\omega\lesssim T_K$) for the Kondo
effect are negligibly small compared to the Fermi energy, the
electronic dispersion relation $\xi_k$ in Eqs.~(\ref{2.2}),(\ref{2.3})
can be linearized: $\xi_k = v_F k$, where $k$ is measured from the
Fermi momentum $k_F$. The linearization leads to an energy-independent
density of states $\nu$, which will be assumed throughout this
paper. Finally, we treat the tunneling amplitudes $t_2,t_R,t_L$ as
real numbers and neglect their dependences on $k$.  This is well
justified for relevant values of $k$, $|k|\lesssim T/v_F$.

Instead of working with the operators $c_{R,L}$, it is convenient to 
introduce their linear combinations $\psi_{0,1}$, 
\begin{equation}
\left( 
\begin{array}{c}
\psi_{1ks} \\
\psi_{0ks}
\end{array}
\right)
=
\left( 
\begin{array}{cc}
\cos\theta_0 & \sin\theta_0 \\
-\sin\theta_0 & \cos\theta_0
\end{array}
\right)
\left( 
\begin{array}{c}
c_{Rks} \\
c_{Lks}
\end{array}
\right) ,
\label{2.5}
\end{equation}
where the angle $\theta_0$ is determined by the equation
\begin{equation}
\tan\theta_0 = t_L/t_R \; . 
\label{2.6}
\end{equation}
(So far there are no restrictions on the value of $t_L/t_R$.) The
Hamiltonian~(\ref{2.1})-(\ref{2.4}) then assumes the ``block-diagonal"
form
\begin{eqnarray}
H &=& H_0\{\psi_0\} + H_1\{\psi_{1},\psi_2, d\} ,
\label{2.7} \\
H_0 &=& \sum_{ks} \xi_k^\pdag \psi^\dagger_{0ks}\psi^\pdag_{0ks},
\label{2.8} \\
H_1 &=&
H_d\{\psi_2,d\} + \sum_{ks} \xi_k^\pdag \psi^\dagger_{1ks}\psi^\pdag_{1ks}  
\label{2.9} \\
&&~~~~~~~ 
+  \sum_{ks}(t_1^\pdag \psi^\dagger_{1ks}d^\pdag_s
+ {\rm H.c.} ) ,
\nonumber
\end{eqnarray} 
where $H_d\{\psi_2,d\}$ is given by Eq.~(\ref{2.2}), and $t_1=
\sqrt{t_L^2 +t_R^2}$.

At low energies ($T\ll E_{1,2}$) the Hamiltonian $H_1$ involving the
$\psi_1$ and $\psi_2$ operators, see Eq.~(\ref{2.9}), can be simplified 
further.  Indeed, at $N\approx 1$ the small dot is occupied by a single 
electron, and, therefore, carries a spin $S=1/2$. The tunneling terms in 
Eqs.~(\ref{2.2}) and (\ref{2.9}) mix the states with a single electron in
in dot $1$ with states having $0$ or $2$ electrons in that dot. Because 
of the high energy cost ($\sim E_1$), these transitions are virtual, and, 
provided that the conductances of the corresponding junctions are small, 
can be taken into account perturbatively in the second order in tunneling 
amplitudes. A new~\cite{OGG} and important element here compared to 
the conventional treatment of the Anderson impurity model is that at $T\ll E_2$ 
only those excitations that conserve the number of electrons in dot $2$ are 
allowed. The resulting effective Hamiltonian which acts within the strip of 
energies $|\omega|\lesssim \min\{E_1,E_2\}$, has the form of the 2CK 
model~\cite{NB,Bethe,AJ,Tsvelik,AL,CZ},
\begin{equation}
H_{2CK} = \sum_{\gamma ks}\xi_k^\pdag 
\psi^\dagger_{\gamma ks} \psi^\pdag_{\gamma ks}
+\sum_\gamma  J_\gamma\left({\bf s}_\gamma\cdot{\bf S}\right) + BS^z.
\label{2.10}
\end{equation}
Here the channel index $\gamma=1$ and $\gamma=2$ represents the leads
and dot $2$, respectively, $\bf S$ is the spin-1/2-operator describing
the doubly-degenerate ground state of dot $1$,
\[
{\bf s}_\gamma = \sum_{kk'ss'} 
\psi^\dagger_{\gamma ks} \frac{{\bm\sigma}_{ss'}}{2}\psi^\pdag_{\gamma k's'}
\]
is the spin density in channel $\gamma$, and $\bm\sigma
=(\sigma^x,\sigma^y,\sigma^z)$ are the Pauli matrices.
The exchange amplitudes $J_\gamma$ in Eq.~(\ref{2.10}) are estimated as
\begin{equation}
\nu J_\gamma = 4\nu t_\gamma^2/E_1.
\label{2.11}
\end{equation}
 
In derivation of Eq.~(\ref{2.10}), we assumed that the gate voltage is
tuned precisely to $N=1$ (which corresponds to a particle-hole
symmetric situation). As we discuss in Section~\ref{Sec4} below, this
assumption does not lead to qualitative changes in the results. We
also included in the Hamiltonian the effect of an external magnetic
field (hereinafter we omit the Bohr magneton $\mu_B$; the field $B$ is
measured in the units of energy).

\section{Tunneling conductance}
\label{Sec33}

In order to study the out-of-equilibrium transport across the device 
we add to our Hamiltonian a term 
\begin{equation}
H_V = \frac{eV}{2} \left(\hat N_L - \hat N_R \right),
\quad
\hat N_\alpha =\sum_{ks} c_{\alpha ks}^\dagger c_{\alpha ks}^\pdag ,
\label{3.1}
\end{equation}
which describes a finite bias voltage $V$ applied between the left
($\alpha=L$) and right ($\alpha=R$) electrodes. The differential conductance $dI/dV$
can be evaluated in a closed form for arbitrary $V$ when one of the
leads, say $L$, serves as a weakly coupled probe~\cite{tutorial}, 
i.e., $t_L\ll t_R$.  Under this condition the angle $\theta_0$ in
Eqs.~(\ref{2.5}) and (\ref{2.6}) is small:
\begin{equation}
\theta_0 \approx t_L/t_R \ll 1.
\label{3.2}
\end{equation}
Application of the transformation Eq.~(\ref{2.5}) to Eq.~(\ref{3.1}),
 yields, to the linear order in $\theta_0$,
\begin{equation}
H_V = \frac{eV}{2}\left(\hat N_0 - \hat N_1\right) 
+ eV \theta_0 \sum_{ks} (\psi^\dagger_{0ks} \psi^\pdag_{1ks} + {\rm H.c.}),
\label{3.3}
\end{equation}
where 
\[
\hat N_0 = \sum_{ks} \psi^\dagger_{0 ks}\psi^\pdag_{0 ks},
\quad
\hat N_1 = \sum_{ks} \psi^\dagger_{1 ks}\psi^\pdag_{1 ks}.
\]
The first term on the right-hand side of Eq.~(\ref{3.3}) can be interpreted 
as a voltage bias between the reservoirs of $0$- and $1$-particles,
cf.~Eq.~(\ref{3.1}), while the second term has an appearance of the 
$k$-conserving tunneling.  Since the tunneling amplitude is proportional 
to the small parameter $\theta_0\ll 1$, see Eq.~(\ref{3.2}), one can use 
perturbation theory to calculate the current across the device~\cite{tutorial}.

Similar to the representation of $H_V$ in the form of Eq.~(\ref{3.3}),
the current operator
\[
\hat I =\frac{d}{dt} \frac{e}{2} \left(\hat N_R - \hat N_L \right) 
\]
also splits naturally into two contributions,
\begin{equation}
\hat I = \hat I_0 + \delta \hat I .
\label{3.4}
\end{equation}
Here 
\begin{eqnarray}
\hat I_0 &=& \frac{d}{dt} \frac{e}{2} \left(\hat N_1 - \hat N_0 \right) 
\label{3.5} \\
&=& ie^2 V\theta_0\sum_{ks} \psi_{0ks}^\dagger \psi_{1ks}^\pdag + {\rm H.c.},
\nonumber
\end{eqnarray}
is a current between the reservoirs of $0$- and $1$-particles, and
\begin{equation}
\delta \hat I = -e\theta_0\frac{d}{dt} 
\sum_{ks}\psi^\dagger_{0ks} \psi^\pdag_{1ks} + {\rm H.c.}
\label{3.6}
\end{equation}
It is easy to show~\cite{tutorial} that in the leading (second) order
in $\theta_0$ the operator $\delta \hat I$ does not contribute to the
average current across the device.  The remaining contribution
$\langle\hat I_0\rangle$ corresponds to the $k$-conserving tunneling
between two bulk reservoirs containing $0$- and $1$-particles, see
Eqs.~(\ref{3.3}) and (\ref{3.5}). Its evaluation
yields~\cite{tutorial}
\begin{equation}
\frac{dI}{dV} = G_0 \sum_s \frac{1}{2} \int d\omega (-df/d\omega) 
\left[-\pi\nu {\rm Im} T_{1s} (\omega +eV)\right]
\label{3.7}
\end{equation}
for the differential conductance. Here $f(\omega)$ is the Fermi function 
($\omega$ is the energy measured from the Fermi level), 
\begin{equation}
G_0 = \frac{2e^2}{h} (2\theta_0)^2 \approx \frac{8e^2}{h} \frac{t_L^2}{t_R^2} ,
\label{3.8}
\end{equation}
and $T_{1s}$ is the t-matrix for the particles of channel $\gamma =1$ 
[evaluated with the equilibrium Hamiltonians 
Eq.~(\ref{2.9}) or Eq.~(\ref{2.10})].  
The t-matrix is related to the exact retarded Green function 
$G_{ks,k's'} = \delta_{ss'} G_{ks,k's}$ 
of these particles according to
\[
G_{ks,k's} = G^0_k  + G^0_k T_{1s}^{\phantom{0}} G^0_{k'},
\quad
G^0_k = (\omega-\xi_k +i0)^{-1}.
\] 
Here we took into account the conservation of the total spin, which
implies that $G_{ks,k's'}$ is diagonal in $s,s'$. In our model with
$t_1$ independent of $k$ (and, consequently, $J_1$ independent of $k$
and $k'$), the t-matrix is also independent of $k,k'$.
Note that the linear response ($V\to 0$) counterpart of Eq.~(\ref{3.7}), 
the linear conductance
\begin{equation}
G = G_0 \sum_s \frac{1}{2} \int d\omega (-df/d\omega) 
\left[-\pi\nu {\rm Im} T_{1s} (\omega)\right],
\label{3.25}
\end{equation}
remains valid~\cite{tutorial} for an arbitrary relation between $t_L$ 
and $t_R$, in which case $G_0 = (2e^2/h)\sin^2(2\theta_0)$.

\section{Transport at finite temperature and bias}
\label{Sec3}

Equation~(\ref{3.7}) provides a direct link between the measurable 
quantity, the differential conductance $dI/dV$, and the properties of 
the 2CK model, Eq.~(\ref{2.10}).  In the channel-symmetric
case $J_1 = J_2=J$ the NFL behavior manifests itself in a nonanalytic
dependence of the t-matrix on energy and temperature~\cite{AL}, which
leads to a rather unusual scaling of the differential conductance at
low bias and temperature ($|eV|,T\ll T_K$):
\begin{equation}
\frac{1}{G_0}\frac{dI}{dV} 
= \frac{1}{2} \left[1 - \sqrt{\frac{\pi T}{T_K}} F_{\rm 2CK}\left(\frac{|eV|}{\pi T}\right)\right].
\label{3.9}
\end{equation}
The function $F_{\rm 2CK}(x)$ here is a universal (parameter-free) scaling
function~\cite{AL} with the asymptotes
\begin{equation}
F_{\rm 2CK}(x) = 
\left\{
\begin{array}{lr}
1+ c x^2, & x\ll 1\\
\displaystyle\frac{3}{\sqrt \pi} \sqrt{x},  & x\gg 1
\end{array}
\right. ,
\label{3.10}
\end{equation}
where $c$ is a numerical coefficient of the order of $1$.  
The limit $eV/T\to 0$ of Eq.~(\ref{3.9}) yields 
\begin{equation}
G = G_0 \frac{1}{2} \left(1 - \sqrt{\pi T/T_K}\right)
\label{3.12}
\end{equation}
for the linear conductance (this result is valid for arbitrary value of $t_L/t_R$). 
The estimate~\cite{scales} of the Kondo temperature $T_K$ 
introduced in Eqs.~(\ref{3.9}) and (\ref{3.12}) reads~\cite{CZ}
\begin{equation}
T_K \sim E_0 (\nu J) e^{-1/J \nu}, 
\qquad
E_0 = \min\{E_1,E_2\}.
\label{3.13}
\end{equation}

\begin{figure}[h]
\includegraphics[width=0.9\columnwidth]{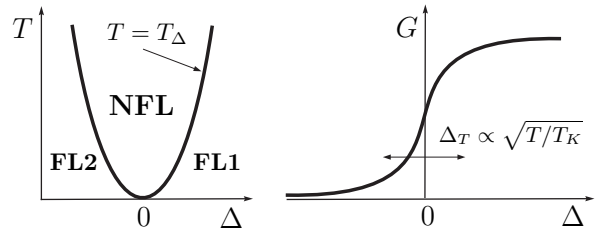} 
\caption{Quantum phase transition between two FL states. The NFL
behavior is preserved at $|\Delta|\neq 0$, provided the temperature
exceeds the crossover scale $T_\Delta$, see Eq.~(\ref{3.18}). The
width $\Delta_T$ of step in the conductance $G(\Delta)$ scales with 
temperature as $\sqrt{T}$, see Eq.~(\ref{3.21}).}
\label{qpt}
\end{figure}

The validity of Eqs.~(\ref{3.9}) and (\ref{3.12}) is limited by the
requirements that both the Zeeman energy $B$ and the level spacing
$\delta E_2$ are small compared to $T$, and that the exchange
constants in Eq.~(\ref{2.10}) are equal to each other: $J_1=J_2$. 
When the system is tuned away from this special point, at a finite
\begin{equation}
\Delta =\nu J_1 - \nu J_2,
\label{3.26}
\end{equation}
the conductance changes drastically.  In the ideal case of $T=0$ and
$\delta E_2 =0$, the conductance has a step-like dependence on $\Delta$,
\begin{equation}
G(\Delta) = G_0 \theta(\Delta).
\label{3.14}
\end{equation}
The discontinuity in Eq.~(\ref{3.14}) reflects a {\it quantum phase
transition} between two different Fermi liquid (FL) states, in which the
spin of the dot $1$ forms a singlet with either the collective spin of
the electrons in the leads (FL1, $\Delta > 0$) or with that of the dot $2$
(FL2, $\Delta < 0$).  At the critical point $\Delta =0$, the system exhibits
NFL behavior down to $T=0$. In agreement with the general theory of
quantum phase transitions~\cite{Sachdev}, the $T\to 0$ asymptotics at
$|\Delta|\neq 0$ corresponds to the FL, whereas the NFL behavior
Eq.~(\ref{3.12}) is preserved at temperatures well above certain 
$\Delta$-dependent crossover scale $T_\Delta$, see Fig.~\ref{qpt}.  
By the same token, the step in the $\Delta$-dependence of $G(\Delta)$, 
Eq.~(\ref{3.14}), is smeared at finite temperatures. 

In order to estimate~\cite{scales} the energy scale $T_\Delta$ we consider 
the renormalization group (RG) flow of the effective exchange constants 
as the high-energy cutoff $D$ is reduced from its initial value $D_0\sim E_0$. 
We are interested in the case when the bare value of $\Delta$ is small,
\[
|\Delta|\ll {\cal J},
\]
where
\begin{equation}
{\cal J} = \nu (J_1+J_2)/2\,.
\label{3.15}
\end{equation}
The evolution of the effective coupling constants ${\cal J}^*, \Delta^*$ 
with the decrease of $D$ is then described by the Poor Man's scaling 
equations~\cite{PWA_book}
\begin{equation}
\frac{d{\cal J}^*}{d\zeta} = ({\cal J}^*)^2, 
\quad
\frac{d\Delta^*}{d\zeta} = 2{\cal J}^* \Delta^*,
\quad
\zeta = \ln\frac{D_0}{D} 
\label{3.22}
\end{equation}
with the initial conditions 
\[
{\cal J}^*(D_0) = {\cal J},
\quad
\Delta^*(D_0) = \Delta.
\] 
Equations~(\ref{3.22}) are valid as long as $\Delta^*\ll {\cal J}^*\ll 1$ and 
yield the relation $\Delta^*/\Delta = ({\cal J}^*/{\cal J})^2$. 
By the time ${\cal J}^*$ has grown to be of the order of $1$ at $D\sim T_K$, 
the value of $\Delta^*$ characterizing the channel asymmetry reaches
\begin{equation}
\Delta^*(T_K) \sim \Delta/{\cal J}^2.
\label{3.16}
\end{equation}
This can be viewed as the initial (at $D\sim T_K$) value of the coupling 
constant of the relevant~\cite{NB,ALPC} 
channel-symmetry-breaking perturbation. The perturbation will eventually 
drive the system away from the 2CK fixed point at $D\to 0$. However, 
if $\Delta^*(T_K)\ll 1$, then one expects the behavior of the system in 
a broad range of energies to be still governed by the vicinity of the 2CK 
fixed point. The channel anisotropy is a relevant operator with scaling 
dimension $1/2$, see Ref.~\onlinecite{ALPC}. Hence, the dependence 
of the corresponding coupling constant $\Delta^*$ on $D$ is described by
\begin{equation}
{ \Delta^*(D) \over \Delta^*(T_K)} \propto \left({ T_K \over D }\right)^{1/2}.
\label{3.17}
\end{equation}
The condition $\Delta^*(T_\Delta) \sim 1$, together with Eq.~(\ref{3.16}),  
then gives the estimate
\begin{equation}
T_\Delta \sim [\Delta^*(T_K)]^2 T_K 
\sim \left({\Delta^2}/{{\cal J}^4}\right) T_K.
\label{3.18}
\end{equation}

\begin{figure}[h]
\includegraphics[width=0.85\columnwidth]{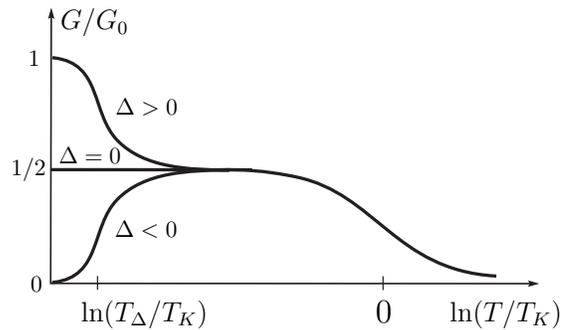} 
\caption{Sketch of the temperature dependence of the linear conductance 
at fixed values of $\Delta$ and $T_K$. For $\Delta<0$ the dependence is 
nonmonotonic, with a maximum at $T\sim \sqrt{T_\Delta T_K}$. At 
$T\gg T_K$ the conductance scales as $G/G_0\propto [\ln(T/T_K)]^{-2}$,
see, e.g., Ref.~\onlinecite{tutorial}.
}\label{G(T)}
\end{figure}

The RG flow stops at $D\sim \max\{T, |eV|\}$. Consequently, at 
$\max\{T_\Delta,|eV|\}\ll T\ll T_K$, 
the channel asymmetry yields a small correction to the conductance 
Eq.~(\ref{3.12}). The correction is first order in the corresponding perturbation,
hence proportional to $\Delta^*(T)\sim (T_\Delta/T)^{1/2}$, 
and its sign is determined by the sign of $\Delta$:
\begin{equation}
\delta G/G_0 \propto \sign (\Delta) \left(\frac{T_\Delta}{T}\right)^{1/2}. 
\label{3.19}
\end{equation}
On the other hand, for $T,|eV|\ll T_\Delta$ the system is a Fermi liquid, see Fig.~\ref{qpt}. 
Substitution of the t-matrix in the form 
\[
-\pi\nu {\rm Im}~ T_{1s} 
= \theta(\Delta) -\sign(\Delta)\frac{3\omega^2 +\pi^2T^2}{2T_\Delta^2}
\]
[cf. Ref.~\onlinecite{AL}] into Eq.~(\ref{3.7}) then yields
\begin{equation}
\frac{1}{G_0}\frac{dI}{dV}= \theta(\Delta) - \sign (\Delta)\left(\frac{\pi
    T}{T_\Delta}\right)^2
\left[ 1 + \frac{3}{2}\left(\frac{eV}{\pi T}\right)^2\right].   
\label{3.20}
\end{equation}
Again, the linear response ($V\to 0$) counterpart of Eq.~(\ref{3.20}) is 
valid at any ratio $t_L/t_R$. The temperature dependence of the linear 
conductance at fixed small values of $\Delta$ is sketched in Fig.~\ref{G(T)}.

According to Eq.~(\ref{3.20}), corrections to the zero-temperature
limit of the linear conductance, the step-function Eq.~(\ref{3.14}), are
quadratic in temperature -- a typical Fermi-liquid result~\cite{Nozieres}. 
At a finite temperature, the step-function is smeared, see Fig.~\ref{qpt}. 
The characteristic width $\Delta_T$ of the smeared step at temperature $T$ 
is estimated by solving the equation $T_\Delta \sim T$ for $\Delta$, 
which results in
\begin{equation}
\Delta_T \sim {\cal J}^2 \sqrt{T/T_K}.
\label{3.21}
\end{equation}
This ``sharpening" of the $\Delta$-dependence of the linear conductance 
with decreasing temperatures (see Fig.~\ref{qpt}) can be regarded as a 
``smoking gun'' for non-Fermi-liquid behavior. In fact, it might be easiest 
to first identify unambiguously the step-like dependence of the conductance
on $\Delta$ and then use it to tune the device precisely to the symmetry point 
in order to observe the distinctive scaling of the differential conductance 
Eq.~(\ref{3.9}).  Experimentally, the value of $\Delta$ is controlled~\cite{OGG}  
by the asymmetry of the conductances of the corresponding tunneling junctions, 
which in turn are controlled by the potentials $V_g$ on the gates forming the
junctions. In the vicinity of the symmetry point, the dependence of $G$ on 
$V_g$ should have the form of a smeared step-function, whose width 
$\delta V_g$ should scale with temperature as $\sqrt T$, see Fig.~\ref{qpt}.

\section{Linear conductance at a finite magnetic field}
\label{Sec4}

The magnetic field dependence of the linear conductance across the 
device also reveals the critical behavior. In this Section we study the 
dependence $G(B)$ at $T=0$ in the vicinity of the quantum critical 
point $\Delta = 0$.  We consider only the Zeeman effect of the 
magnetic field, and dispense with its orbital effect (this is an adequate 
approximation for a field applied in the plane of a lateral quantum dot 
device).

Similar to the effect of a finite temperature, see Fig.~\ref{qpt}, the
application of a magnetic field at small $\Delta$ results in a crossover 
from the limiting FL behavior at $B\to 0$ to NFL intermediate regime 
at higher fields $B\gtrsim B_\Delta$.  As before, the crossover scale 
$B_\Delta$ can be estimated~\cite{scales} from RG arguments. The 
scaling dimension~\cite{ALPC} of the operator $S^z$ in Eq.~(\ref{2.10}) 
at the 2CK fixed point is $1/2$. Accordingly, when the high energy cutoff 
$D$ is lowered, the effective splitting of the impurity levels $B^*$ evolves 
according to
\begin{equation}
{ B^*(D)/D \over  B^*(T_K) / T_K} 
 \propto\left(\frac{T_K}{D}\right)^{1/2}
\label{4.6}
\end{equation}
with the initial condition $B^*(T_K)\sim B$. The RG flow Eq.~(\ref{4.6}) 
terminates once $B^*$ has grown to become of the order of $D$, or 
when $D$ reaches the value $T_\Delta$, whichever occurs at a higher value 
of $D$.  The first of the two conditions corresponds to the limitation on the 
NFL behavior set by the Zeeman splitting, while the second one is due to
the channel anisotropy. Therefore, the crossover scale $B_\Delta$ can be 
estimated as that field $B \sim B^*(T_K)$ in Eq.~(\ref{4.6}), at which 
$B^*(D)\sim D$ and $D\sim T_\Delta$ simultaneously. Using Eqs.~(\ref{4.6}) 
and (\ref{3.18}), we find the relation between the crossover field~\cite{AJ}, 
the crossover temperature $T_\Delta$, and the channel anisotropy parameter 
$\Delta$
\begin{equation}
B_\Delta
\sim \sqrt{T_\Delta T_K}
\sim \left({|\Delta|}/{\cal J}^2\right)T_K .
\label{4.7}
\end{equation}
Note the difference between the $\Delta$--dependence of the crossover
temperature $T_\Delta$ [Eq.~(\ref{3.18})] and the crossover field $B_\Delta$. 

Having found the crossover scale $B_\Delta$, next we investigate
the dependence of the conductance $G$ on the field $B$. First of all,
we note that at $\Delta\neq 0$ the low-energy properties of the 
Hamiltonian Eq.~(\ref{2.10}) are those of a Fermi liquid.\cite{NB}  
The effect of any \textit{local} perturbation, such as the exchange 
interaction with the spin of the dot $1$ in Eq.~(\ref{2.10}),
on the ground state of the Fermi liquid is completely characterized by
the scattering phase shifts $\delta_{\gamma s}$ at the Fermi level.
(Recall that $s=\pm 1$ for spin-up/down and $\gamma =1,2$ labels the
two channels.)  The t-matrix that enters Eq.~(\ref{3.25}) is then given
by the standard scattering theory expression
\begin{equation}
-\pi\nu T_{\gamma s}(0) = \frac{1}{2i} 
\left(e^{2i\delta_{\gamma s}} - 1\right) .
\label{4.1}
\end{equation}
Obviously, the phase shifts are defined only mod $\pi$ (that is,
$\delta_{\gamma s}$ is equivalent to $\delta_{\gamma s}+\pi$ ). The
ambiguity is removed by setting the values of the phase shifts
corresponding to $J_\gamma = 0$ in Eq.~(\ref{2.10}) to zero. With this
convention, the invariance of the Hamiltonian~(\ref{2.10}) with
respect to the particle-hole transformation $\psi_{\gamma k s}\to
\psi^\dagger_{\gamma, -k,- s}$ translates into the relation
\begin{equation}
\delta_{\gamma s} + \delta_{\gamma, -s} = 0 
\label{4.2}
\end{equation}
for the phase shifts, which suggests a representation
\begin{equation}
\delta_{\gamma s} = s \delta_{\gamma} .
\label{4.3}
\end{equation}
Substitution of Eqs.~(\ref{4.1}) and (\ref{4.3}) into Eq.~(\ref{3.25}) yields 
\begin{equation}
G/G_0 = \frac{1}{2}\sum_s \sin^2\delta_{1s}
= \sin^2\delta_1
\label{4.4}
\end{equation}
for the linear conductance at $T=0$. In the limit $B/T_K\to +0$ and at
$\Delta\neq 0$, the ground state of the Hamiltonian~(\ref{2.10})
is a singlet.  Therefore, the total spin in a very large but finite
region of space surrounding the dot $1$ is zero.  By the Friedel sum
rule, this implies relation $\sum_{\gamma s} s\delta_{\gamma s} =
\pi$.  Taking, in addition, Eq.~(\ref{4.3}) into account, one obtains
relation
\begin{equation}
\delta_1 +\delta_2 = \pi/2,
\label{4.5}
\end{equation}
valid at any value of $B/B_\Delta$, as long as $B\ll T_K$.

Below the crossover, $B\ll B_\Delta$, the values of the phase
shifts are determined by the vicinity of the stable Fermi-liquid fixed
points~\cite{NB}, $\delta_1 =\pi/2$, $\delta_2=0$ at $\Delta >0$ and
$\delta_1 = 0$, $\delta_2=\pi/2$ at $\Delta <0$.  Substitution of
these values into Eq.~(\ref{4.4}) then yields Eq.~(\ref{3.14})
for the conductance. The corrections to the fixed point values of the
phase shifts are linear in $B/B_\Delta$,
\begin{equation}
\delta_1 = \pi/2 -\delta_2 
= (\pi/2)\theta(\Delta) - \sign (\Delta) (B/B_\Delta),
\label{4.8}
\end{equation}
yielding
\begin{equation}
G/G_0 = \theta(\Delta) - \sign (\Delta) (B/B_\Delta)^2 ,
\quad
B\ll B_\Delta
\label{4.9}
\end{equation}
[cf. Eq.~(\ref{3.20})].

Above the crossover, i.e., for $B_\Delta\ll B\ll T_K$, the departure 
of the phase shifts from the 2CK fixed point values $\delta_{1,2}=\pi/4$ 
is controlled by the properties of the fixed point. To account for a 
finite value of $B/T_K$, we generalize Eq.~(\ref{4.5}):
\[
\delta_1 + \delta_2 = \pi\left[1/2 + M(B)\right].
\] 
The zero-temperature magnetization $M(B)$ here is known exactly from the
Bethe-ansatz solution~\cite{Bethe,AJ,SS}. Using the asymptote~\cite{SS}
$M(B)\propto (B/T_K)\ln (T_K/B)$, we find
\begin{equation}
\delta_1 = { \pi \over 4} + \,a\,\sign(\Delta) {B_\Delta \over B} - b
{B \over T_K} \ln {T_K \over B} .
\label{4.10}
\end{equation}
Here $a$ and $b$ are positive numerical coefficients of the
order of $1$.  The second term on the right-hand side of Eq.~(\ref{4.8}) is the
first-order correction in the channel-symmetry-breaking perturbation.
This correction is similar to Eq.~(\ref{3.19}) with temperature $T$
replaced by the energy scale $D^*(B)\sim B^2/T_K$ at which the RG flow
defined by Eq.~(\ref{4.6}) terminates.  Equations~(\ref{4.10}) and
(\ref{4.4}) yield the asymptote of the conductance at $B_\Delta\ll
B\ll T_K$,
\begin{equation}
\frac{G}{G_0} = \frac{1}{2} + \,a\,\sign(\Delta)\frac{B_\Delta}{B} 
- b \frac{B}{T_K}\ln\frac{T_K}{B} .
\label{4.11}
\end{equation}
The shape of $G(B)$ is qualitatively similar to that of $G(T)$, see 
Eqs.~(\ref{3.12}),(\ref{3.19}), and (\ref{3.20}), although the 
precise functional form is rather different. 

Interestingly, in the case of small channel anisotropy, $T_\Delta\ll T_K$,
there is an approximate symmetry with respect to the change of sign of 
$\Delta$:
\begin{equation}
G(B,\Delta) + G(B,-\Delta) = 2G(B,\Delta\to 0).
\label{4.12}
\end{equation}
Note that this relation is valid at any $B/T_K$, provided that
$T_\Delta / T_K \ll 1$.

Strictly speaking, the consideration of this Section is applicable only 
at zero temperature. However, the results Eqs.~(\ref{4.9}) and (\ref{4.11}) 
remain valid~\cite{CZ} as long as 
\begin{equation}
T \ll B^2/T_K 
\label{4.13}
\end{equation}
At higher temperatures the conductance is described by the corresponding 
expressions of Section~\ref{Sec3}. As follows from Eqs.~(\ref{3.12}) 
and (\ref{4.11}), the limiting value of the linear conductance at the 2CK 
fixed point, $G=G_0/2$, is independent of the order in which the limits 
$B\to 0$, $T\to 0$ are taken~\cite{nCK}. Hence, the crossover between 
the field-dominated regime, see Eqs.~(\ref{4.9}) and (\ref{4.11}), and 
the temperature-dominated one, see Eqs.~(\ref{3.12}), (\ref{3.19}) and 
(\ref{3.20}), is expected to be smooth and featureless.

For arbitrary values of $T_\Delta/T_K$, the detailed magnetic field
dependence of the phase shifts at the Fermi level can be studied using
the numerical renormalization group (NRG)~\cite{Wilson}.  In this
approach one defines a sequence of discretized Hamiltonians and
diagonalizes them iteratively to obtain the finite-size spectrum of
the model.  In the Fermi liquid case ($\Delta\neq 0$) knowledge of the
finite-size spectrum is sufficient to identify unambiguously the phase
shifts~\cite{ALPC}.

\begin{figure}[h]
\includegraphics[width=0.75\columnwidth]{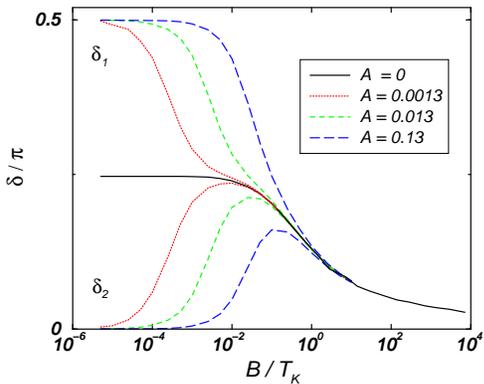}
\caption{The phase shifts for the 2CK model at different values of the channel 
asymmetry parameter $A= \Delta/{\cal J}^2$.  The upper (lower)
curves represent $\delta_1(\delta_2)$.  
}
\label{NRG1}
\end{figure}

\begin{figure}[h]
\includegraphics[width=0.9\columnwidth]{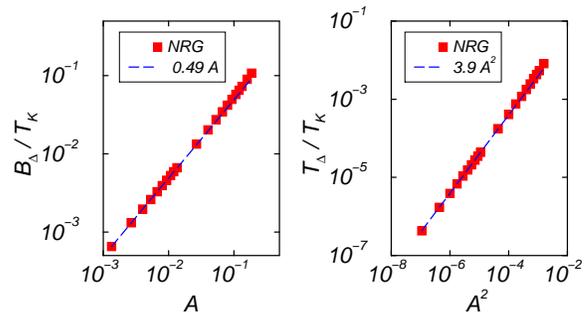}
\caption{Dependences of the crossover scales $B_\Delta$ and $T_\Delta$ 
on the asymmetry parameter $A= \Delta/{\cal J}^2$.  
}
\label{NRG2}
\end{figure}

In Fig.~\ref{NRG1}, we plotted the phase shifts $\delta_{1,2}$ as a
function of $B$ for different values of the parameter 
$A= \Delta/{\cal J}^2>0$ that characterizes the asymmetry between 
the channels. We estimate the crossover scales~\cite{scales} $T_K$ 
and $B_\Delta$ as the two values of $B$ in  Fig.~\ref{NRG1} at 
which the phase shift $\delta_2$ equals $\pi/8$. In order to verify 
the relation $B_\Delta/T_K\sim A$, see Eq.~(\ref{4.7}), we plotted
$B_\Delta$ vs $A$ on the left panel in Fig.~\ref{NRG2}. The NRG 
data also allow us to estimate the scale $T_\Delta$, see Eq.~(\ref{3.18}), 
as the energy scale at which the first excited state of the NRG spectrum 
has reached the halfway mark of its crossover evolution between the 
corresponding two fixed point values, see Fig.~\ref{NRG2}, right panel.
The NRG data are very well described by $B_\Delta / T_K \approx 0.5 A$, 
$T_\Delta / T_K \approx 4 A^2$, in agreement with Eqs.~(\ref{4.7}) and 
(\ref{3.18}) above.

\begin{figure}[h]
\includegraphics[width=0.75\columnwidth]{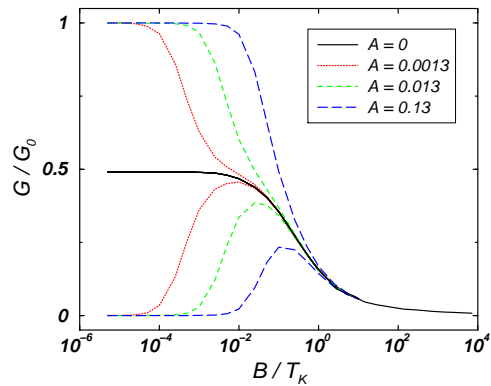}
\caption{Field dependence of the conductance at different values of the 
asymmetry parameter $A=\Delta/{\cal J}^2$. The upper (lower) curves 
correspond to $A>0$ ($A<0$).
}
\label{NRG3}
\end{figure}

Having extracted the phase shifts, we are able to calculate the linear
conductance from Eqs.~(\ref{4.4}) and (\ref{4.12}), see
Fig.~\ref{NRG3}.  As expected, the conductance develops a signature of
a plateau at intermediate values of the field $B_\Delta<B<T_K$. At
very high fields, $B\gg T_K$, the conductance scales with $B$ as
$1/\ln^{2}(B/T_K)$.

As usual in NRG calculations, we measured all energies in units 
of the bandwith $D$. In order to avoid the disturbing finite 
bandwidth effects, we used two different coupling constants for the 
high- and low-field regimes: one set of data, that
includes the $B\gg B_\Delta$ regime, was obtained using 
${\cal J}=0.075$, while  another set of data, which includes the 
$B \ll T_K$ regime, was obtained using ${\cal J}=0.15$. 
The two sets were combined by rescaling the magnetic field in units of
the Kondo temperature, resulting in a set of continuous
curves, as shown in the figures. The overlap of the two sets of data
at intermediate fields confirms that in this regime the accuracy 
of our numerics is remarkably good. Based on the dependence on 
the finite system size, we estimate the relative error of the calculated 
phase shifts to be of the order of $2\%$. (The worst case is the low 
field part of the $A=0$ curve, because of the extremely fragile nature 
of the intermediate NFL fixed point.) 

\section{Effect of Potential Scattering}
\label{Sec5}

So far we concentrated on the particle-hole symmetric model. In
general, however, such symmetry is absent. It is violated by the
presence of higher energy levels in dot $1$, and also by deviations
of the dimensionless gate voltage $N$ from an integer value. In the
absence of particle-hole symmetry, the effective Hamiltonian~(\ref{2.10}) 
acquires additional terms leading to potential scattering. Taking into 
account that the interchannel scattering is blocked at energies well below 
$E_{1,2}$, we can write this additional perturbation as
\begin{equation}
H_p = \sum_{\gamma=1,2} V_\gamma 
\sum_{k k' s}\psi^\dagger_{\gamma k s} \psi^\pdag_{\gamma k' s} .
\label{5.1}
\end{equation}
Including $H_p$ into our considerations leads to a modification of the
limiting values of the conductance in the Fermi-liquid and 2CK fixed
points. The dependences of $dI/dV$ on $\Delta,V,T$ and $B$, however,
remain the same apart from acquiring a constant background
contribution $G_{el}$ due to elastic cotunneling. Here we illustrate
this for a specific example of the zero-temperature magnetoconductance.

The potential scattering yields finite spin-independent phase shifts
$\delta_\gamma^0 = -\arctan(\pi\nu V_\gamma)$ even if $J_\gamma$ in
Eq.~(\ref{2.10}) are set to $0$.  This can be accounted for by a
proper modification~\cite{N} of Eq.~(\ref{4.3}),
\begin{equation}
\delta_{\gamma s} = \delta_\gamma^0 + s\delta_\gamma, 
\label{5.2}
\end{equation}
where the dependence of $\delta_\gamma$ on $B$ and $\Delta$ is
described by the "particle-hole symmetric" expressions~(\ref{4.8}) and
(\ref{4.10}). Substitution of the phase shifts in the form of Eq.~(\ref{5.2}) 
into Eq.~(\ref{4.4}) results in~\cite{real}
\begin{equation}
G(B,\Delta) = G_{el} + {\widetilde G}_0 F[B/B_\Delta, B/T_K,\sign(\Delta)],
\label{5.3}
\end{equation}
where $G_{el} = G_0\sin^2\delta^0_1$, the function $F$ is a universal
function with asymptotes given in Eqs.~(\ref{4.9}) and (\ref{4.11}), and
${\widetilde G}_0 = G_0 - 2G_{el}$. Note that the limiting value of
the conductance at the 2CK fixed point, $G_{el} + {\widetilde G}_0/2$, 
lies precisely {\it half-way} between the two Fermi-liquid limits,
$G_{el}$ and $G_{el} + {\widetilde G}_0$, and that Eq.~(\ref{4.12})
remains valid even in the presence of the potential scattering
Eq.~(\ref{5.1}).

\section{Discussion}

The low-temperature properties of a quantum dot device normally are
well described by  Fermi liquid theory. The special two-dot
structure proposed in Ref.~\onlinecite{OGG} allows, however, for NFL
behavior at a special point in the space of parameters of the
device. In the context of the physics of quantum phase transitions, this
point can be viewed as a critical point separating two Fermi liquid
states. In this paper, we developed a detailed theory of the transport
properties near such a quantum critical point. Our theory offers a
strategy for tuning the device parameters to the critical point
characterized by the two-channel Kondo effect physics, by monitoring
the temperature dependence of the linear conductance, see
Sec.~\ref{Sec3}. Further confirmation of the 2CK behavior may come
from the measurements of the differential conductance, which must
display universal behavior, see Sec.~\ref{Sec3}. We also
investigated the effect of magnetic field and of potential scattering
on the conductance in the vicinity of the quantum critical point, see
Secs.~\ref{Sec4} and \ref{Sec5}. The Zeeman splitting allows one to
investigate the finite-field crossover between the Fermi liquid and
NFL behavior of the conductance. In the vicinity of the NFL point, the
linear conductance of the device depends on the magnetic field and
temperature only via two dimensionless parameters, $T/T_\Delta$ and
$B/B_\Delta$; the dependence of $T_\Delta$ and $B_\Delta$ on the
channel asymmetry $\Delta$ is given in Eqs.~(\ref{3.18}) and
(\ref{4.7}). Note also that potential scattering does not destroy
the 2CK behavior, but merely renormalizes the magnitude of the Kondo
contribution to the conductance. A finite level spacing in the larger
dot $\delta E_2$, however, is a hazard. At temperatures below $\delta E_2$
the two-dot device inevitably enters into the conventional
Fermi-liquid regime.

\bigskip
\begin{acknowledgments}
We are grateful to the Aspen Center for Physics, Max Planck Institute
for the Physics of Complex Systems (Dresden), and LMU M\"unchen for
hospitality and thank N. Andrei, A. Ludwig, Y. Oreg, A. Rosch, A. Tsvelik, 
and G. Zar\'and for discussions. The research at the University of Minnesota 
was supported by NSF grants DMR02-37296 and EIA02-10736. L.B. 
acknowledges the financial support provided through the European 
Community's Research Training Networks Programme under contract 
HPRN-CT-2002-00302, Spintronics. 
\end{acknowledgments}


\begin{thebibliography}{99}
\bibitem{scales} The definitions of the crossover scales $T_K$, 
$T_\Delta$ and $B_\Delta$ adopted in this paper are based on the 
asymptotic behavior of the linear conductance at low $T$ and $B$, 
see Eqs.~(\ref{3.12}), (\ref{3.20}) and (\ref{4.9}), correspondingly.
\bibitem{nCK}
Note that if the number of channels is larger than $2$, the value of the 
conductance at $T\to 0$ and $B\to 0$ depends on the order in which the limits 
are taken. 
This property is easy to understand in the limit of very large number of channels 
when the NFL fixed point lies within the reach of perturbation theory~\cite{NB,AL,FR}.
\end{thebibliography}
\end{document}